\documentclass[english,10pt, onecolumn, notitlepage,  superscriptaddress]{revtex4-1}
\usepackage[T1]{fontenc}
\usepackage[latin9]{inputenc}
\usepackage{amstext}
\usepackage{graphicx}

\def\7{$\;$}
\def\l{\left}
\def\r{\right}
\def\be{\begin{equation}}
\def\ee{\end{equation}}
\def\bea{\begin{eqnarray}}
\def\eea{\end{eqnarray}}
\def\f{\frac}

\def\k{\kappa}

\def\no{\nonumber}
\def\d{{\rm d}}

\usepackage{babel}
\def\be{\begin{equation}}
\def\ee{\end{equation}}
\def\bea{\begin{eqnarray}}
\def\eea{\end{eqnarray}}
\def\no{\nonumber}
\def\f{\frac}
\def\l{\left}
\def\r{\right}

\makeatother

\usepackage{babel}
\begin{document}

\title{Cosmology of non-minimal derivative coupling to gravity in Palatini formalism and its chaotic inflation}

\author{Narakorn Kaewkhao}\email{narakornk56@email.nu.ac.th}
\affiliation{The Institute for Fundamental Study ``The Tah Poe Academia Institute'', Naresuan University, Phitsanulok 65000, Thailand}

\author{Burin Gumjudpai}\email{buring@nu.ac.th: corresponding}
\affiliation{The Institute for Fundamental Study ``The Tah Poe Academia Institute'', Naresuan University, Phitsanulok 65000, Thailand}

\date{\today}
\begin{abstract}

We consider, in Palatini formalism, a modified gravity of which the scalar field derivative couples to Einstein tensor. In this scenario, Ricci scalar, Ricci tensor and Einstein tensor are functions of connection field. As a result, the connection field gives rise to relation, $h_{\mu\nu} = f g_{\mu\nu}$ between effective metric, $h_{\mu\nu}$ and the usual metric $g_{\mu\nu}$ where $f \,=\,1 - \kappa{\phi}^{,\alpha}{\phi}_{,\alpha}/2 $.  In FLRW universe, NMDC coupling constant is limited in a range
of $ -2/ \dot{\phi}^{2} <   \kappa        \leq  \infty $ preserving Lorentz signature of the effective metric.  Slowly-rolling regime provides $\kappa < 0$ forbidding graviton from travelling at superluminal speed. Effective gravitational coupling and entropy of blackhole's apparent horizon are derived.
 In case of negative coupling, acceleration could happen even with $w_{\rm eff} > -1/3$. Power-law potentials of chaotic inflation are considered.   For $V \propto \phi^2$ and $V \propto \phi^4$, it is possible to obtain tensor-to-scalar ratio lower than that of GR so that it satisfies $r < 0.12$ as constrained by Planck 2015 \cite{Ade:2015lrj}.  The $V \propto \phi^2$  case yields acceptable range of  spectrum index and $r$ values. The quartic potential's spectrum index is disfavored by the Planck results. Viable range of $\k$ for $V \propto \phi^2$  case lies in positive region, resulting in less blackhole's entropy, superluminal metric, more amount of inflation, avoidance of super-Planckian field initial value and stronger gravitational constant.

\end{abstract}

\maketitle

\section{Introduction}
\label{Section:Introduction}
 Astrophysical observations strongly convinces us that the space is in the state of accelerating expansion.
 Results obtained from supernova type Ia (SNIa)
 \cite{Amanullah2010, Astier:2005qq, Goldhaber:2001a, Perlmutter:1997zf, Perlmutter:1999a, Riess:1998cb, Riess:1999ar, RiessGold2004, Riess:2007a, Tonry:2003a},
 large-scale structure surveys \cite{Scranton:2003, Tegmark:2004}, cosmic microwave background (CMB) anisotropies
 \cite{Larson:2010gs, arXiv:1001.4538, CMBXRay:2014, Masi:2002hp} and X-ray luminosity from galaxy clusters \cite{CMBXRay:2014, Allen:2004cd, Rapetti:2005a} are
 examples of the evidence of the acceleration.
 If the expansion is to be accelerated, some unknown form of dark energy \cite{Copeland:2006a, Padmanabhan:2004av, Padmanabhan:2006a}
 is suggested as  a driving force of the dynamics.
Typically dark energy is in form of cosmological constant or scalar field  \cite{Copeland:2006a, Padmanabhan:2004av, Padmanabhan:2006a, AT:DE2010}
 such as quintessence \cite{Caldwell:1997ii}-scalar with canonical kinetic term, and classes of k-essence type kinetic energy \cite{Chiba:1999ka, ArmendarizPicon:2000dh, ArmendarizPicon:2000ah} which are hypothesized as dark energy.  Alternative of Einstein gravity such as  braneworlds, $f(R)$ could as well result in present acceleration (see e.g.  \cite{AntoShijiLRR2010, Carroll2004, Nojiri:2010wj}). Other the other situation, inflationary expansion \cite{Guth, Sato, Starobinsky, Linde, AbSh} in the early universe is also strongly supported by the recent CMB anisotropy observations \cite{Ade:2015lrj}. Scalar field models or modified gravities should provide explanation to either or both inflationary acceleration and present acceleration.

A gravitational theory with non-minimal coupling (NMC) between scalar field's derivative term and a gravity sector could
as well give accelerating expansion.
In metric formalism, of which the metric $g_{\mu\nu}$ is a dynamical variable, the coupling function $f(\phi, \phi_{,\mu}, \phi_{,\mu\nu}, \ldots)$ can be motivated
by requirement of scalar quantum electrodynamics to preserve U(1) symmetry
or by models with gravitational constant as a function of the mass density \cite{Amendola1993}.
Non-minimal derivative coupling (NMDC) to $R$ term
 can be  found in lower energy limits of higher dimensional
theories and in Weyl anomaly of $N = 4$ disformal supergravity
\cite{Liu:1998bu, Nojiri:1998dh}.
Without loss of generality,  other possible coupling terms apart from $R \phi_{,\mu}\phi^{,\mu}$
and  $R^{\mu\nu} \phi_{,\mu} \phi_{,\nu}$ are not necessary  \cite{Capozziello:1999xt}. Hence gravitational theory with NMDC terms,
 $R^{\mu\nu} \phi_{,\mu} \phi_{,\nu}$ and $R \phi_{,\mu}\phi^{,\mu}$ with a free canonical kinetic term but without  $V(\phi)$ nor $\Lambda$ term
 was studied and found cosmologically interesting, i.e. it gives de Sitter expansion  \cite{Capozziello:1999uwa}. Moreover types of  NMDC models with two separated couplings have been investigated in various contexts and with further modifications  \cite{Granda:2010hb, Granda:2010ex, Granda:2011zk}.

Considering  a special case of $\kappa_1 R \phi_{,\mu}\phi^{,\mu}$ and $\kappa_2 R^{\mu\nu} \phi_{,\mu} \phi_{,\nu}$  term,
one can set $\kappa \equiv   \kappa_2  =  -2 \kappa_1$. As a result, the two NMDC terms combine into the Einstein tensor, $G^{\mu\nu}$ coupling to scalar field's kinetic part as
$\kappa G_{\mu \nu} \phi^{,\mu}\phi^{,\nu}$. The field equations contain terms with second-order derivative in $g_{\mu\nu}$ and $\phi$ at most order
hence it is a good dynamical theory as Lagrangian contains only divergence-free tensors  \cite{Sushkov:2009}.
In flat FLRW universe, for $\kappa > 0$, there is quasi-de Sitter phase at very early stage and there is initial singularity at very early stage for $\kappa < 0$.
The expansion is $a \propto t^{1/3}$ at very late time for any sign of $\kappa$ \cite{Sushkov:2009}.
When adding $V = \text{constant}$ and allowing phantom sign of the free kinetic term, it is possible to
transit from de-Sitter phase to other types of expansions  \cite{Saridakis:2010mf}.
If without free kinetic term, the model gives superluminal sound speed  \cite{Gao:2010vr} and if having both free and coupling
$\kappa G_{\mu \nu} \phi^{,\mu}\phi^{,\nu}$ term with $V(\phi) = 0$, the model can not have phantom crossing.
Therefore potential is added into the theory and the acceptable action is
\be\label{MainAction}
S = \int {\d}^4x {\sqrt{-g}}\l[\f{R}{8\pi G_{\rm N} }- \l(\varepsilon g_{\mu\nu} + \kappa
G_{\mu\nu}\r)\phi^{,\mu}\phi^{,\nu} - 2V(\phi)\r] + {S}_{\rm m}\,.   \label{jjj}
\ee
It is found that the potential must be less steep than quadratic potential in acquiring inflation \cite{Skugoreva:2013ooa}.
With a constant potential and a matter term in the model, it is able to
describe transition from inflation to matter domination epoch without reheating and this description includes transition to late de-Sitter epoch \cite{Sushkov:2012}.
For positive potential, $\kappa > 0$ gives unbound $\dot{\phi}$ value with restricted Hubble parameter \cite{Skugoreva:2013ooa}. For
 $V = \text{constant}$ and $\kappa > 0$, inflationary phase is always possible and the inflation depends solely on the value of coupling constant.
 Gravitational heavy particles are less produced during inflation when coupling to the inflaton field or to the particles  gets stronger \cite{Koutsoumbas:2013boa}.
Perturbations and inflationary analysis of the model with a cosmological constant (or equivalently, the constant potential) was shown in
 \cite{Darabi:2013caa} confronting observational data and in
 \cite{Dalianis:2016wpu} as of exponential and monomial potentials. Other very interesting studies of slightly different versions of NMDC model are reported, see such as
\cite{Germani:2009, Germani:2010gm, Germani:2011ua, Sadjadi:2012zp,Tsujikawa:2012mk, Ema:2015oaa,Ghalee:2014bta,Yang:2015pga,Myung:2015tga,
Sadjadi:2010bz, Sadjadi:2013psa, Gumjudpai:2015vio, Sheikhahmadi:2016wyz, Gumjudpai:2016frh}.  The NMDC model considered here falls into a subclass of the Horndeski action (with $G_5 = \phi \kappa/2 $) which is generalized action that is Ostrogradski instability free \cite{Horn}.

So far the results given above are obtained in metric formalism. Considering Einstein-Hilbert action with matter term, the metric formalism gives equivalent field equation as
that of the Palatini formalism. When GR is modified, the Palatini approach does not give the same field equations as those of the metric formalism
as there is a non-minimal coupling between geometrical part and matter field and/or having some form of functions of the Ricci scalar.
The affine connection and the metric are fundamentally independent concepts of geometrical entities \cite{Koba}.
Hence in the Palatini formalism, we consider metric tensor and connection field
as independent dynamical fields.
This independent connection is not the usual Levi-Civita connection constructed from the metric $g_{\mu\nu}$ \cite{M, pp1, pp2, Capoz} but another
independent field which does not couple to matter fields (see e.g. \cite{Olmo:2011uz, Iglesias:2007nv, Clifton:2011jh} for a review on Palatini approach and its theoretical motivations).
The Palatini approach to the NMC model was investigated before in \cite{Allemandi:2005qs}.
Recently the Palatini approach to the NMDC model with two separated coupling constants, i.e. one of the $R$ coupling term
and the other one of the $R_{\mu \nu}$ term with a non-zero potential, have been reported by some authors \cite{palaNMDC}.  It was found that phantom crossing with oscillating equation of state parameter is possible. In this work, the NMDC term is in form of one Einstein tensor coupling to the scalar field's kinetic term. Hence there is
only one combined coupling constant as in Eq. (\ref{jjj}) and we will treat the action (\ref{jjj}) with the Palatini formalism.
The relation between the metric and effective metric of the connection field may look like disformal type \cite{Capozziello dark metric:2008} (generalization introduced by Bekenstien \cite{Bekenstein:1992}). However here the situation is not about transformation between conformal or disformal frames. Not to mislead the readers, therefore we shall not refer to it
as conformal nor disformal factor, but only a factor or a relation.

We shall investigate cosmological scenario of the model in Sec. \ref{sec_cosmo}. whereas it is suggested that the field should be slowly-rolling and the coupling constant should be negative
in order to preserve Lorentz invariance and to prevent the graviton from traveling faster than light. Cosmological field equations under slow-roll condition are stated in Sec. \ref{sec_srr}. Inflationary consideration is explored in Sec. \ref{sec_sri} in which slow-roll parameters and spectral index were derived. We consider power-law potential of chaotic inflation in Sec. \ref{sec_sripower}. We conclude our work and give comments in Sec. \ref{sec_con}.

\section {Palatini NMDC gravity}    \label{sec_PNMDC}
In the metric formalism, Sushkov's NMDC action is \cite{Sushkov:2009}
\bea  \label{action NMDC metric}
S_{g} &=& \int {\d^4 x } {\sqrt{-g}} \l\{  {R(g)} - \l[\varepsilon g_{\mu\nu}+ \kappa\l(R_{\mu\nu}(g)-\frac{1}{2}  g_{\mu\nu}R(g)\r)\r] \phi^{,\mu}\phi^{,\nu}-2V (\phi) \r\} +{S}_{\rm m}[g_{\mu\nu},\Psi],
\eea
where the $\Psi$ represents matter fields and we set the unit $c = 1$ and $8 \pi G_{\rm N} = 1$.
Canonical scalar field Lagrangian density is
$\mathcal{L}_{\phi}=-\varepsilon g_{\mu\nu}\phi^{,\mu}\phi^{,\nu} -2V(\phi)$ where $\varepsilon =\pm1$ is for canonical and the phantom cases respectively.
The Einstein tensor is conventional,
$
 G_{\mu\nu}(g)=R_{\mu\nu}(g)-\frac{1}{2}g_{\mu\nu}R(g)\,,
$
as in Eq. (\ref{jjj}). Differently, in Palatini formalism, the NMDC action is expressed as \cite{palaNMDC}
\bea\label{action NMDC Palatini}
{S}_{\rm Palatini} &=&\int{\d^4 x }{\sqrt{-g}} \l\{  {\tilde{{R}}(\Gamma)}  -  \l[\varepsilon g_{\mu\nu}+ \kappa_{1}
g_{\mu\nu}\tilde{{R}}(\Gamma)+\kappa_{2}\tilde{{R}}_{\mu\nu}(\Gamma)\r]\phi^{,\mu}\phi^{,\nu}-2V(\phi)\r\}+ {S}_{\rm m}[g_{\mu\nu},\Psi]\,.
\eea
Tilde symbol denotes variables that depend on the connection field.
Following Sushkov  \cite{Sushkov:2009},
we set  $\kappa=\kappa_{2}=-2\kappa_{1}$ and
define the Einstein tensor in Palatini formalism,
\be\label{Einstein field in Palatini}
\tilde{G}_{\mu\nu}(\Gamma)=\tilde{R}_{\mu\nu}(\Gamma)-\frac{1}{2}g_{\mu\nu}\tilde{R}(\Gamma).
\ee
Hence Eq. (\ref{action NMDC Palatini}) is
\bea\label{action NMDC Palatini2}
S_{\rm Palatini} &=
&\int{\d^4x }{\sqrt{-g}} \l\{   {\tilde{{R}}(\Gamma)}
-    \l[     \varepsilon   g_{\mu\nu}      +         \kappa\tilde{{G}}_{\mu\nu}(\Gamma)                            \r]        \phi^{,\mu}\phi^{,\nu}-2V(\phi)\r\} +   {S}_{\rm m}[g_{\mu\nu},\Psi].
\eea
The Palatini Ricci tensor is defined by the independent dynamical connection,
\be\label{Ricci tensor pala nmdc}
               \tilde{{R}}_{\mu\nu}(\Gamma)\;=\;{\tilde{{R}}^{\lambda}}_{\,\mu\lambda\nu}(\Gamma)\;=\;\partial_{\lambda}{{\Gamma}^{\lambda}}_{\,\mu\nu}-\partial_{\nu}{{\Gamma}^{\lambda}}_{\,\mu\lambda}
               +{{\Gamma}^{\lambda}}_{\,\sigma\lambda}{{\Gamma}^{\sigma}}_{\,\mu\nu}-{{\Gamma}^{\lambda}}_{\,\sigma\nu}{{\Gamma}^{\sigma}}_{\,\mu\lambda},
\ee
and the Palatini Ricci scalar is  $\tilde{R}=\tilde{{R}}(\Gamma)=g^{\mu\nu}\tilde{{R}}_{\mu\nu}(\Gamma)$.
Varying the Palatini NMDC action in Eq. (\ref{action NMDC Palatini2}) with respect to the metric, we obtain the first field equation,
\bea \label{vary g pala NMDC}
 T_{\mu\nu}
&=&
\tilde{G}_{\mu\nu}(\Gamma)+ \Bigg[\frac{\kappa}{2}\tilde{G}_{\mu\nu}(\Gamma)\phi_{,\lambda}\phi^{,\lambda}+\frac{\kappa}{2}\tilde{R}_{\alpha\beta}(\Gamma)g_{\mu\nu}\phi^{,\alpha}\phi^{,\beta}
-\kappa \tilde{R}_{\nu\lambda}(\Gamma)\phi_{,\mu}\phi^{,\lambda}+\frac{\kappa}{2}\tilde{R}(\Gamma)\phi_{,\mu}\phi_{,\nu}\no \\&&-2\kappa \tilde{R}_{\mu\lambda}(\Gamma)\phi_{,\nu}\phi^{,\lambda}+\frac{\varepsilon}{2}g_{\mu\nu}\phi_{,\alpha}\phi^{,\alpha}
-\varepsilon\phi_{,\mu}\phi_{,\nu}+g_{\mu\nu}V(\phi)\Bigg]\,,
\eea
whereas the matter energy-momentum tensor is
\be\label{energy momentum tensor}
T_{\mu\nu}=-\frac{2}{\sqrt{-g}}\frac{\delta S_{\rm m}[ g_{\kappa\lambda},\Psi]}{\delta g^{\mu\nu}}.
\ee
The second Palatini NMDC field equation comes from the second degree of freedom, the independent connection field ${{\Gamma}^{\lambda}}_{\mu\nu}$ and it is
\be\label{vary gamma nmdc4}
{\nabla}^{\Gamma}_{\lambda}\l\{\sqrt{-g}\l[g^{\mu\nu}\l(1-\frac{1}{2}\kappa{\phi}^{,\alpha}{\phi}_{,\alpha}\r)\r]\r\}=0,
\ee
where $\nabla^{\Gamma}_{\lambda}$ is the covariant derivative with respect to the independent connection. This is written as
\be\label{vary gamma nmdc5}
{\nabla}^{\Gamma}_{\lambda}\l(\sqrt{-g}g^{\mu\nu}f\r)\:=\:0,
\ee
where
\be\label{read f f dot}
f \,=\,1-\frac{1}{2}\kappa{\phi}^{,\alpha}{\phi}_{,\alpha}.
\ee
Solving Eq. (\ref{vary gamma nmdc5}), the new metric or the effective metric $h_{\mu\nu}$ is related to the metric $g_{\mu\nu}$ via a transformation factor $f$ as
\be\label{new metric nmdc}
h_{\mu\nu} =f g_{\mu\nu}=(1-\frac{1}{2}\kappa{\phi}^{,\alpha}{\phi}_{,\alpha})g_{\mu\nu}\,,  \label{eq_Disss}
\ee
of which $\sqrt{-h}=\sqrt{-g}f^2$ and its inverse is $
h^{\mu\nu} = f^{-1} g^{\mu\nu}.
$
Conformal invariance and disformal invariance between dual con(dis)formal frames \cite{Capozziello dark metric:2008, Allemandi:2004, Santos:2012, Bekenstein:1992, Zumalacarregui:2010, Shinji:2015} are not the case here since
the $h_{\mu \nu}$ metric is the effect of the Palatini connection field, not mathematical transformation of the Lagrangian from one frame to another.
The relation is in form of
\be\label{disformal transformation}
{h}_{\mu\nu}\equiv\alpha(\phi,X)g_{\mu\nu}+\beta(\phi,X)\phi_{,\mu}\phi_{,\nu},
\ee
where $\alpha(\phi,X)$ and $\beta(\phi,X)$ are generalized factors. In general, the factors
depend on the field kinetic term, $X=g^{\mu\nu}\nabla_{\mu}\phi\nabla_{\nu} \phi$.
The effective metric, $h_{\mu\nu}$  in Eq. (\ref{new metric nmdc}) is hence related to $g_{\mu\nu}$ with $\beta(\phi,X) = 0$.
 Eq. (\ref{disformal transformation}) can be written as
\bea\label{disformal in general}
h_{\mu\nu}  =  \alpha(\phi,X)g_{\mu\nu}  
= \l[\alpha_{1}(\phi)+\alpha_{2}(\phi)\phi^{,\sigma}\phi_{,\sigma}\r]g_{\mu\nu}\,,
\eea
such that   $\alpha_{1}(\phi)=1$ and $\alpha_{2}(\phi)={-\kappa}/{2}$.
The relation (\ref{eq_Disss}) allows us to write the action (\ref{action NMDC Palatini2}) as function of  $h_{\mu\nu}$,
\be\label{action conformal}
{S}_{\rm Palatini}  =
\int \d^4 x {\sqrt{-h}}   \l\{ \f{\tilde{R}(h)}{f^2}  -    \l[\frac{\varepsilon h_{\mu\nu}}{f^3}    + {\kappa} \f{\tilde{G}_{\mu\nu}(h)}{f^2}
             \r] \phi^{,\mu}\phi^{,\nu}  -    \f{2 {V(\phi)}}{f^2} \r\} + S_{\rm m} \l( \frac{h_{\mu\nu}}{f}, \Psi \r)\,.
\ee

\section {Cosmological scenario}   \label{sec_cosmo}
The factor (\ref{disformal in general})  in flat FLRW geometry with homogenous scalar field is
\be\label{read f f dot2}
f(\dot{\phi}) \; =\; 1-  \frac{\kappa}{2} g^{00} \frac{\d \phi}{\d t}{\frac{\d \phi}{\d t}}\;=\;1+\frac{\kappa}{2}{\dot{\phi}}^2\,.
\ee
 Note that there is a factor $8 \pi G_{\rm N} \equiv 1$ multiplying with $\kappa$  in this equation.   With Eq. (\ref{disformal in general}), the new metric $h_{\mu\nu}$ preserves Lorentz signature (-,+,+,+)  if  $\alpha_2(\phi) \,=\, {-\kappa}/{2} $  and       $     -2/ \dot{\phi}^{2} <   \kappa    $  due to positivity of  the  factor
 $f(\dot{\phi})$.
For fast-rolling field, the coupling is allowed in positive region or in very small negative-value region. Hence
the coupling estimably ranges from $0 < \kappa$.
For slowly-rolling field,  the coupling is permitted in vast negative region.
The FLRW effective metric  is therefore
\be\label{new metric NMDC}
h_{\mu\nu}=\left(
  \begin{array}{cccc}
    -1-\frac{\kappa}{2}{\dot{\phi}}^2  & 0 & 0 & 0 \\
    0 & a^2(1+\frac{\kappa}{2}{\dot{\phi}}^2) & 0 & 0 \\
    0 &0 & a^2(1+\frac{\kappa}{2}{\dot{\phi}}^2) & 0 \\
    0 & 0 & 0 & a^2(1+\frac{\kappa}{2}{\dot{\phi}}^2) \\
  \end{array}
\right)\,, \ee
of which graviton speed is modified with the Palatini NMDC effect.  Slowly-rolling field allows negative $\kappa$ hence graviton travels under the speed of light. On the other hand,  fast-rolling case could result in superluminal graviton. We hence restrict our consideration to the slowly-rolling case.
The result above enables us to find that
$\nabla^{\Gamma}_{\lambda}(\sqrt{-h}h^{\mu\nu}) = \nabla^{\Gamma}_{\lambda}(\sqrt{-g}g^{\mu\nu}f)$ and the
independent connection $\Gamma^{\lambda}_{\,\mu\nu}(h)$ is constructed with the effective metric $h_{\mu\nu}$ as
\be
\label{h form connection}
\Gamma^{\lambda}_{\,\mu\nu}(h)=\frac{1}{2}h^{\lambda\sigma}\l(\partial_{\mu}
h_{\sigma\nu}+\partial_{\nu}h_{\sigma\mu}-\partial_{\sigma} h_{\mu\nu} \r).
\ee
Following e.g. \cite{Brustein:2009, Faraoni:2011hf}, the effective gravitational coupling of Palatini NMDC gravity is hence
\be
G_{\rm eff} = \f{f^{2}}{8 \pi} = \f{1}{8 \pi}{\l(1+\frac{\kappa}{2}{\dot{\phi}}^2 \r)^{2}}, \ee
leading to modification of the entropy of blackhole's apparent horizon for this theory,
\be S_{\rm AH} = \f{A}{{4 (1+\frac{\kappa}{2}{\dot{\phi}}^2)^2}/ 8 \pi}. \ee
The effective gravitational coupling strength of the model could be tested by observing
temporal variation of the effective gravitational coupling,
\be\label{ratio Geff NMDC}
\frac{\dot{G}_{\rm eff}}{G_{\rm eff}}=\frac{2\kappa \dot{\phi}{\ddot{\phi}}}{(1+\frac{\kappa}{2}\dot{\phi}^2)}.
\ee
For fast-rolling field, $
{\dot{G}_{\rm eff}}/{G_{\rm eff}}  \simeq   {4{\ddot{\phi}}}/{\dot{\phi}}\,
$ and on the other hand, $
{\dot{G}_{\rm eff}}/{G_{\rm eff}}  \simeq  2\kappa{\dot{\phi}}{\ddot{\phi}}\,
$ for slowly-rolling field.
Since $\Gamma = \Gamma(h, \partial h) $ hence the field equation (\ref{vary g pala NMDC}) is expressed
as function of $h$ and $ \partial h$, for example,
$\tilde{R}_{\mu\nu}(\Gamma)$ is  $ \tilde{R}_{\mu\nu}(h, \partial h)$ (written $\tilde{R}_{\mu\nu}(h)$ for brevity).
The energy-momentum tensor obeys the relation  $\tilde{T}_{\mu\nu}=f^{-1}T_{\mu\nu}$ (see Appendix \ref{sec_ap}).
Considering time-time component of the field equation,
\be\label{T00 NMDC}
{T}_{00}=\tilde{G}_{00}(h) -\frac{\kappa}{2}{\tilde{G}}_{00}(h){\dot{\phi}}^2 +\frac{5\kappa}{2}\tilde{R}_{00}(h){\dot{\phi}}^2+\frac{\kappa}{2}\tilde{R}(h){\dot{\phi}}^2-\Big(\frac{\varepsilon}{2}{\dot{\phi}}^2+V(\phi)\Big).
\ee
The Ricci tensor for the effective metric $h_{\mu\nu}$ in $n$ dimensions is related to the usual Ricci tensor by the following formula (see e.g. \cite{maeda, Carroll Book}),
\be\label{Ricci tensor consformal tranf}
\tilde{R}_{\sigma\nu}(h) = {R}_ {\sigma\nu}(g)-\l[(n-2)\delta^{\alpha}_{\sigma}\delta^{\beta}_{\nu}+g_{\sigma\nu}g^{\alpha\beta}\r] \frac{1}{\sqrt f}(\nabla^{g}_{\alpha}\nabla^{g}_{\beta}\sqrt{f})
+\l[2(n-2)\delta^{\alpha}_{\sigma}\delta^{\beta}_{\nu}-(n-3)g_{\sigma\nu}g^{\alpha\beta}\r] \frac{1}{f}(\nabla^{g}_{\alpha}\sqrt{f})(\nabla^{g}_{\beta}\sqrt{f}),
\ee
where $\nabla^{g}_{\lambda}$ is the  usual covariant derivative constructed from $g_{\mu\nu}$.
Ricci tensor and Ricci scalar for the metric $g_{\mu\nu}$ in flat FLRW universe are,
\be\label{FRLW Ricci tensor and scalar}
{R}_{00}(g)=-3(\dot{H}+H^2),\qquad {R}_{ii}(g)=a^2(\dot{H}+3H^2),\qquad {R}(g)=6(\dot{H}+2H^2)\,,
\ee
where $H = {\dot{a}}/{a}$.
In case of the metric $h_{\mu\nu}$
\be
\tilde{R}_{00}(h)
=-3(\dot{H}+H^2) - \frac{3}{2}\l(\frac{\ddot{f}}{f}-\frac{{2\dot{f}}^2}{f^2}\r)\,, \qquad
\tilde{R}_{ii}(h)  =   R_{ii}(g) + \frac{a^2\ddot{f}}{2f}.  \label{R11 h NMDC}
\ee
First and second-order time derivative of the factor are expressed in term of $\dot{\phi},\ddot{\phi}$ and $\dddot{\phi}$, i.e.
\be\label{phi dot phe dd dot}
\dot{f}\,=\, \kappa\dot{\phi}\,\ddot{\phi}\,,   \qquad  \qquad         \ddot{f}\,=\, \kappa\l({\ddot{\phi}}^2+\dot{\phi} \;  \dddot{\phi} \r).
\ee
The Ricci scalar is (see \cite{Carroll Book})
\be\label{Ricci scalar conformal NMDC}
\tilde{R}(h) \:=\: f^{-1}{R}(g)-2(n-1)g^{\alpha\beta}f^{-3/2}\l(\nabla_{\alpha}\nabla_{\beta}{\sqrt{f}}\r)-(n-1)(n-4)g^{\alpha\beta}f^{-2}\l(\nabla_{\alpha}\sqrt{f}\r)\l(\nabla_{\beta}\sqrt{f}\r).
\ee
In four  dimensions, this is
\be\label{Ricci scalar conformal NMDC2}
\tilde{R}(h) \; = \;   \frac{1}{f}{R}(g)+3\l(\frac{\ddot{f}}{f^2}-\frac{{\dot{f}}^2}{2f^3}\r) \;=  \;   \frac{6}{f}\l(\dot{H}+2H^2\r)+3\l(\frac{\ddot{f}}{f^2}-\frac{{\dot{f}}^2}{2f^3}\r).
\ee
Using $\tilde{T}_{\mu\nu} = f^{-1}T_{\mu\nu}$, the $T_{00}$ component of NMDC-Palatini field equation is the matter density,
\be
 \rho_{\rm m} =
\dot{H}\Big[12f+\frac{6}{f}-18\Big]+H^2\Big[12f+\frac{12}{f}-21\Big]
- \frac{3}{2} (1-f)\Big(\frac{4\ddot{f}}{f}-\frac{8\dot{f}^2}{f^2}\Big)
- \frac{3\ddot{f}}{2f}+\frac{3\ddot{f}}{f^2}+\frac{3\dot{f}^2}{f^2}-\frac{3\dot{f}^2}{2f^3}
 - \rho_{\phi},
    \label{rho eff NMDC for h metric1}
\ee
where  ${\rho}_{\rm tot}
    \equiv    {\rho_{\rm m}}+ \rho_{\phi}$ and $ \rho_{\phi} =  {\varepsilon {\dot{\phi}}^2}/{2}   +  V(\phi)$.
The matter pressure found from the ${{T}}_{ii}$ components is
\be\label{Pressure eff NMDC for h metric2}
 p_{\rm m}  =
\dot{H}\l(4f-6\r)+H^2\l(6f-9\r)    -\f{3}{2}(1-f)\l(\frac{\ddot{f}}{f}-\frac{2\dot{f}^2}{f^2}\r)
+\frac{\ddot{f}}{f}-\frac{3\ddot{f}}{2f}+\frac{3\dot{f}^2}{4f^2}  -   p_{\phi}\,,
\ee
where $ {p}_{\rm tot}\equiv    p_{\rm m} + p_{\phi}$.
The pressure of scalar field is $p_{\phi}= \varepsilon {\dot{\phi}}^2/{2}-V(\phi)$.
For brevity, we define
\bea\label{define variables}
A& \equiv &4f-6,\\
B& \equiv &  6f-9
,\\ C&  \equiv & -\frac{3}{2}(1-f)(\frac{\ddot{f}}{f}-\frac{2\dot{f}^2}{f^2})+\frac{\ddot{f}}{f}-\frac{3\ddot{f}}{2f} +\frac{3\dot{f}^2}{4f^2} ,\\
D& \equiv &12f+\frac{6}{f}-18,\\
 E& \equiv &  12f+\frac{12}{f}-21,\\
 F & \equiv &-\frac{3}{2}(1-f)(\frac{4\ddot{f}}{f}-\frac{8\dot{f}^2}{f^2})-\frac{3\ddot{f}}{2f}+\frac{3\ddot{f}}{f^2}+\frac{3\dot{f}^2}{f^2}-\frac{3\dot{f}^2}{2f^3}\,.
\eea
The effective equation of state  parameter is hence
\be\label{weff new}
w_{\rm eff}\equiv\frac{{p}_{\rm tot}}{{\rho}_{\rm tot}}=\frac{A\dot{H}+BH^2+C}{D\dot{H}+EH^2+F}\,.
\ee
The modified $\dot{H}$ and modified Friedmann equations are found as (the Friedmann equation in the $h_{\mu\nu}$ metric is shown in the Appendix B.)
\be
\dot{H}=\frac{\l[(B -  Ew_{\rm eff})H^2-Fw_{\rm eff}+C\r]}{Dw_{\rm eff}-A}\,,
\qquad
H^2=
\frac{ {\rho}_{\rm tot}}{3}\frac{\l[1-\frac{(  C-F  w_{\rm eff} )  D}{(D  w_{\rm eff}       -  A)
 {\rho}_{\rm tot}} - \frac{F}{ {\rho}_{\rm tot}}\r]}{\l[{\frac{(B-E w_{\rm eff}  )D}{3(D  w_{\rm eff}    - A)}+\frac{E}{3}}\r]}\,.\label{Hubble NMDC in tern rho and p}
\ee
The acceleration equation can be found from the $ {\ddot{a}}/{a}  =  \dot{H} + H^2 $.
In the GR limit,  $f =1$ making $A=-2$, $B=-3$, $E=3$ and $C=D=F=0$ hence ${p}_{\rm tot}$ and $w_{\rm eff}$ reduce to the usual
$
 {p}_{\rm tot}=-2\dot{H}-3H^2  $ and $
w_{\rm eff}   = -1- {2\dot{H}}/{3H^2}.
$
The Friedmann and acceleration equations can as well reduce to standard GR case,
$
H^2= {\rho}_{\rm tot}/3,
$
and
$
{\ddot{a}}/{a} =
- \l(1/{6}\r)  \l({\rho}_{\rm tot}+3 {p}_{\rm tot}\r).
$
Modified Klein-Gordon equation can be obtained from the Euler-Lagrange equation for scalar field (see e.g. \cite{Carroll Book} for standard result),
\be\label{KG NMDC}
\ddot{{\phi}}\l[-\varepsilon+\frac{\kappa}{2}\l(\tilde{R }(h)-{\tilde{R}_{00}(h)}\r)\r]-\kappa\dot{{\phi}}\nabla_{0}^{h}\tilde{R}_{00}(h)+\frac{\kappa}{2}\dot{{\phi}}\nabla_{0}^{h}\tilde{R}(h)-3\varepsilon H\dot{{\phi}}-V'=0,
\ee
where  $V'={\d V(\phi)}/{\d\phi}$ and $
{\nabla}_{\mu}^h \phi=\nabla_{\mu}^{g}\phi=\partial_{\mu}\phi\,,
$
and
\be\label{conformal scalar second derivative}
{\nabla}_{\mu}^{h}{\nabla}_{\nu}^{h}\phi
=
\nabla_{\mu}^{g}\nabla_{\nu}^{g}\phi - \l(\delta^{\alpha}_{\mu}\delta^{\beta}_{\nu}+\delta^{\beta}_{\mu}\delta^{\alpha}_{\nu}-g_{\mu\nu}g^{\alpha\beta}\r)
 \frac{1}{\sqrt f} \l(\nabla_{\alpha}^{g}{\sqrt f}\r) \l(\nabla_{\beta}^{g} \phi \r).
\ee
The time component of the equation (\ref{conformal scalar second derivative}) reads,
$
{\nabla}_{0}^{h}{\nabla}_{0}^{h}\phi =  \ddot{\tilde{{\phi}}} =      \ddot{\phi}/f\,.
$
The modified Klein-Gordon equation hence reads
\bea\label{modified KG eq2}
\ddot{\phi}
&&\l\{-\varepsilon
+\frac{\kappa}{2}\l[\l(\frac{6\dot{H}+12H^2}{f}+3\l(\frac{\ddot{f}}{f^2}-\frac{{\dot{f}}^2}{2f^3}\r)\r)-\l(-3\l(\dot{H}+H^2\r)-\frac{3}{2}\l(\frac{\ddot{f}}{f}-\frac{2{\dot{f}}^2}{f^2}\r)\r)\r]\r\}
\no \\ &&+\frac{\kappa}{2}\dot{\phi}\nabla_{0}^{g}\l[\frac{6\dot{H}+12H^2}{f}+3\l(\frac{\ddot{f}}{f^2}-\frac{{\dot{f}}^2}{2f^3}\r)\r]-3\varepsilon H\dot{\phi}-V'=0.
\eea
which  recovers the usual Klein-Gordon equation,
$
\varepsilon \ddot{\phi}+3 \varepsilon H\dot{\phi} +V'=0
$
in the GR limit.
\section{slow-roll regime}  \label{sec_srr}

Slowing-rolling field obeys the condition $ 0 <  |\dot{\phi}| \ll 1$ and we approximate further that  $|\dddot{\phi}| \ll  |\ddot{\phi}|   \ll   |\dot{\phi}|$, i.e.  $0 \sim  |\ddot{f}|  \ll  |\dot{f}|  \ll |f|$. This enables us to neglect
$\dddot{\phi}$, $\dot{\phi}^4 \ddot{\phi}^2$,  $\dot{\phi}\, \dddot{\phi}$  terms and to do binomial approximation, $(1+\kappa \dot{\phi}^2/2)^{-1} \simeq 1 - \kappa \dot{\phi}^2/2 $ hence $1/f \simeq 2 - f $ in the calculation. As a result, $
A  \simeq   -2 +2\kappa \dot{\phi}^2,
B  \simeq   -3   +3 \kappa \dot{\phi}^2,
D  \simeq   3 \kappa \dot{\phi}^2,
E  \simeq  3,
C  \simeq  F \simeq 0\,
$
and $A \simeq {2}B/3$.  We found that
\bea\label{accelerate eq NMDC}
\frac{\ddot{a}}{a} &\simeq&  -\f{1}{6} \rho_{\rm tot} \l[  1 +\f{7}{2}{\kappa \dot{\phi}^2}+ 3 w_{\rm eff} \l(1+\f{3}{2}\kappa \dot{\phi}^2\r) \r],
\eea
giving the acceleration condition,
\be
w_{\rm eff}  \,  \lesssim       \,   - \f{1}{3} \l( 1 +2\kappa \dot{\phi}^2 \r).
\ee
which is found from the square bracket of the Eq. (\ref{accelerate eq NMDC}).  As discussed earlier, the slowly-rolling field allows negative $\kappa$ hence the expansion is under acceleration at $w_{\rm eff}$ slightly greater than $-1/3$. On the other hand, if $\k > 0$,  it is needed that $w_{\rm eff}$ is less than $-1/3$ in order to have acceleration.
The modified Friedmann equation (\ref{Hubble NMDC in tern rho and p}) is approximated as
\be
H^2 \; \simeq  \; \f{1}{3} {\rho}_{\rm tot}\l[\frac{3(Dw_{\rm eff}-A)}{BD-EA} \r]  \;
   \simeq \;  \f {{\rho}_{\rm tot}}{3}
  \l[ 1+\frac{3}{2}\kappa \dot{\phi}^2(1+w_{\rm eff} )\r]\,,   \label{Friedmann slowroll}
\ee
and the modified Klein-Gordon equation (\ref{modified KG eq2}) in the slow-roll regime is
\bea\label{MF KG}
\ddot{\phi}\l\{\varepsilon  - \f{9 \kappa}{2} \dot{H}\l(   1  - {\kappa} \dot{\phi}^2     \r)  -  \f{3\kappa}{2}  H^2 \l(   5   - {6 \kappa \dot{\phi}^2 } \r)  \r\}
+ 3 H\dot{\phi}  \l[ \varepsilon   -\l( \f{\ddot{H}}{H}  + 4 \dot{H}       \r)\kappa \l( 1 - \f{\kappa \dot{\phi}^2}{2}  \r)  \r]   + V' \simeq 0.
\eea

\section{Slow-roll inflation}  \label{sec_sri}
Considering the early universe when the scalar field dynamically drives the inflation. Scalar field density dominates the universe and slow-roll regime is plausible during the era.
The Friedmann equation (\ref{Friedmann slowroll}) reads
\be\label{Friedmann Pala infla3}
H^2 \simeq    \f {{\rho}_{\phi}}{3 M^2_{\rm P}}
  \l[ 1+\frac{3}{2}\kappa \dot{\phi}^2 \l(1+\frac{{p_{\phi}}}{{\rho_{\phi}}} \r)\r]
\: \simeq \: \frac{1}{3}\frac{V(\phi)}{M^2_{\rm P}},
\ee
as the condition $\dot{\phi}^2\ll V(\phi)$ is assumed.
Here we restore $8\pi G_{\rm N} = M^{-2}_{\rm P}$ to the equation. The coupling constant can be considered in $mass^{-2}$ dimension, $\kappa=M^{-2}  \lesssim   M^{-2}_{\rm P}.$ The other useful relation is
\be
\dot{H}\,  \simeq \,  \f{V'\dot{\phi}}{6HM^2_{\rm P}}\,  \simeq \,  \f{\sqrt{3} V' \dot{\phi}}{6 \sqrt{V} M_{\rm P}}   
 \label{eq_dotHbaba}
\ee
as the slow-roll approximated Friedmann equation is used.
Further slow-roll approximation, $\ddot{\phi}\dot{\phi}^2 \simeq 0$,  $\dot{\phi}^3 \simeq 0$ and
  $|\ddot{H}|\ll |H\dot{H}| < |H^{3}|$ give the Eq. (\ref{MF KG}) as
\be
\ddot{\phi}\l( \varepsilon - \f{9 \kappa \dot{H}}{2}  -  \f{15 \kappa}{2}H^2 \r)  +  3H \dot{\phi} \l( \varepsilon - 4 \dot{H} \kappa  \r)   + V'  \;  \simeq \;  0\,.  \label{eq_KGpalaq}
\ee
If we let  $\ddot{\phi}\simeq 0$, this gives
\be
\dot{\phi} \, \simeq  \, -\frac{V'(\phi)}{3H(\varepsilon-4\kappa \dot{H})}\,. \label{phi dot inflation}
\ee
Using this in (\ref{eq_dotHbaba}) hence,
\bea\label{H dot substitute}
\dot{H} &\simeq&
-\frac{(V'(\phi))^2}{18H^2M^2_{\rm P}(\varepsilon -4\kappa\,\dot{H})}.
\eea
Deriving slow-roll parameters, the first one is
\bea
\epsilon_{\rm v} \,\equiv\, -\frac{\dot{H}}{H^2} \,&\simeq& \,  \f{M_{\rm P}^2}{2 (\varepsilon -4\kappa\,\dot{H})}        \l(\frac{V'}{V}\r)^2, \label{def of epsilon infla}
\eea
with a slow-roll condition, $\epsilon_{\rm v} \ll 1$.  The second parameter
\bea\label{second slowroll delta}
\delta \, \equiv \,   \frac{\ddot{\phi}}{H\dot{\phi}}
   \, \simeq  \,  -\frac{V''(\phi)}{3H^{2}(\varepsilon-4\kappa \dot{H})}      +  \frac{V'(\phi)\dot{H}}{3H^3(\varepsilon - 4\kappa \dot{H})\dot{\phi}}  -
   \frac{4\kappa \ddot{H}V'(\phi)}{3 M_{\rm P}^2 H^2 (\varepsilon-4\kappa \dot{H})^{2}   \dot{\phi}}\,,
\eea
with a slow-roll condition,
$|\delta|  \ll 1 $. The first term on the right-hand side of Eq.(\ref{second slowroll delta}) is, in fact, the other slow-roll parameter,
\bea
\eta_{\rm v}  &\equiv&       \frac{M^2_{\rm P}}{(\varepsilon-4\kappa \dot{H})}\frac{V''(\phi)}{V(\phi)}                      \simeq   \frac{V''(\phi)}{3H^{2}(\varepsilon-4\kappa \dot{H})}
          \label{eta def}
\eea
with a slow-roll condition, $ |\eta_{\rm v} | \ll 1$.  The second term on the right-hand side of Eq. (\ref{second slowroll delta}) is just
$ \epsilon_{\rm v} $ while the third term, with help of Eq.  (\ref{phi dot inflation}),  is a new slow-roll parameter, $\eta_{\k}$, manifesting another NMDC-Palatini effect,
\bea\label{third term delta}
        \eta_{\k}  \, \equiv  \, \frac{4\kappa \ddot{H}}{H M_{\rm P}^2(\varepsilon-4\kappa \dot{H})}   \,    \simeq  \,  -\frac{4\kappa \ddot{H}V'(\phi)}{3H^2 M_{\rm P}^2 (\varepsilon-4\kappa \dot{H})^{2} \dot{\phi}} \: \simeq\:    \f{4 \k}{M_{\rm P}^2 (\varepsilon  - 4 \k \dot{H})^3} \l[  \f{V'' (V')^2}{18 V}   -  \f{V'^4}{36 V^2 }  \r]\,.
\eea
Therefore $  \delta   =  - \eta_{\rm v} + \epsilon_{\rm v} + \eta_{\k}$.
The spectral index of the model can be derived from
$n_{\rm s}  - 1   =  - 4 \epsilon_{\rm v}  - 2 \delta $  to obtain \cite{Liddle:1994dx},
\be n_{\rm s}  - 1   =  - 6 \epsilon_{\rm v}  + 2 \eta_{\rm v} - 2 \eta_{\k}  \ee
or written in full form of $V(\phi)$ and its field derivatives,
\be
n_{\rm s}-1 =  -\f{3 M_{\rm P}^2}{(\varepsilon-4 \k \dot{H})}\l( \f{V'}{V} \r)^2  +  \f{2 M_{\rm P}^2}{(\varepsilon-4 \k \dot{H})} \f{V''}{V}
  - \f{8 \k}{M_{\rm P}^2 (\varepsilon-4 \k \dot{H})^3}\l[\f{V'' (V')^2}{18 V} - \f{(V')^4}{36 V^2} \r]\,.   \label{eq_nspala}
\ee
The $e$-folding number during the inflationary epoch can be found from
$
\mathcal{N}=\ln \l({a_{\rm f}}/{a_{\rm i}}\r) = \int^{t_{\rm f}}_{t_{\rm i}}H\,\d t,
$
where  ${\rm i}$ and ${\rm f}$ denotes the beginning and the end of inflationary phase.
From Eqs. (\ref{Friedmann Pala infla3})  and (\ref{phi dot inflation}),
$
H \d t =[{-V (\varepsilon-4\kappa \dot{H})}/{(V' M^{2}_{\rm P})}] \d \phi.
$
During inflation, $\dot{H}$ is almost constant due to slowly-rolling field as seen in Eq. (\ref{eq_dotHbaba}), the number of $e$-folds is hence approximately
\bea
\mathcal{N} \; \simeq \;
\frac{(\varepsilon-4\kappa \dot{H})}{M^{2}_{\rm P}}\int^{\phi_{\rm i}}_{\phi_{\rm f}} \frac{V(\phi)}{V'(\phi)}\d\phi\, = \,{(\varepsilon-4\kappa \dot{H})}
 \int^{\phi_{\rm i}}_{\phi_{\rm f}}    \frac{1}{\sqrt{2 \epsilon_{\rm v, GR}}}   \f{\d\phi}{M_{\rm P}}\,,  \label{eq_Nmo}
\eea
where here $\epsilon_{\rm v, GR} \equiv  (M_{\rm P}^2 /2)(V'/V)^2$.  A known from of the scalar field potential is necessary in order to evaluate $n_{\rm s}$ and $\epsilon_{\rm v}$. From now on, we consider only the non-phantom field, i.e. $\varepsilon =1$. For slow-roll scalar field dynamics, $\dot{H} < 0$,
if the NMDC-Palatini effect, $\kappa < 0$, it reduces the amount of inflation from that of the GR case. On the other hand, for $\kappa > 0$, the effect is opposite, that is to enhance the amount of inflation.

\section{Chaotic inflation potentials}   \label{sec_sripower}
Consider single monomial potential (chaotic inflation \cite{Linde1983}) in form of
\be V(\phi)\,=\, V_{0}\phi^{n}\,, \ee
where $V_0 \equiv \lambda(M_{\rm P}^4/M_{\rm P}^n)$. With
${V'}/{V}= {n}/{\phi}\,$  and $
{V''}/{V} = {n(n-1)}/{\phi^2},
$ the slow-roll parameters and the spectral index can be found,
\be
\epsilon_{\rm v}  \, = \,  \f{n^2}{2(1 - 4 \k \dot{H})} \f{M_{\rm P}^2}{\phi^2}  \,,\hspace{0.5cm}    \eta_{\rm v}\, =\, \f{n(n-1)}{(1 - 4 \k \dot{H}) } \f{M_{\rm P}^2}{\phi^2}\,, \hspace{0.5cm}   \eta_{\k}\, =\, \f{n^3 (n-2)}{9(1 - 4 \k \dot{H})^3} \k V_0^2 \f{\phi^{2n - 4}}{M_{\rm P}^2} \ee
\be
n_{\rm s} - 1 \, = \, - \f{M_{\rm P}^2 \l[ n(n+2)  \r]}{( 1 - 4\kappa \dot{H}  )} \,{\phi^{-2}}   \,-\,   \f{2 \k V_0^2 \l[ n^3(n-2)  \r]}{9 M_{\rm P}^2 (1-4\kappa \dot{H})^3}\,  \phi^{2n - 4}\,.
\ee
In GR, scalar field rolling in power-law potential (with $n > \sqrt{2}$) is super-Planckian in order to satisfy the slow-roll condition. Considering $\epsilon_{\rm v} \ll 1$,  the GR slow-roll condition is modified with the NMDC-Palatini effect,
$
[{|n|}/(\sqrt{2} \sqrt{1-4\kappa \dot{H} }) ] M_{\rm P} \:  <  \: \phi\,.
$
For  $\k > 0$ case, the slowly-rolling scalar field can avoid the super-Planckian regime if  the coupling takes the value in a range,
$
\k    <   - ({n^2  - 2})/(8 \dot{H}) \,,
$
(note that $\dot{H} < 0$). That is for $V \propto \phi^{-2}$ or $ V \propto \phi^2$, it is $\k < 1/(4 |\dot{H}|)$ and for $V \propto \phi^4$, it is $\k <  7 /(4 |\dot{H}|)$.  Oppositely if $\k < 0$, the field takes more super-Planckian value in order to slowly roll.
For simplification, we let $\phi_{\rm f} = 0$ and the integral (\ref{eq_Nmo}) is $
\mathcal{N} \; \simeq \; {(1 - 4\kappa \dot{H})}   {\phi^2_{\rm i}}/({2 n M^{2}_{\rm P}})
$ so that we define
\be\label{phi function of N}
\phi^2 \, \equiv \,  \phi_{\rm i}^{2}(n,\mathcal{N}, \dot{H}) \, \simeq \,  \frac{2n M^2_{\rm P}}{(1 - 4\kappa \dot{H})}  \mathcal{N}     \,.
\ee
The $e$-folding number of the GR case is $\mathcal{N}_{\rm GR} = \phi^2/(2 n M_{\rm P}^2)$, therefore $ \mathcal{N} = \mathcal{N}_{\rm GR}  (1 - 4\kappa \dot{H})$ and $\mathcal{N} > \mathcal{N}_{\rm GR}$ for $\k > 0$.
The slow-roll parameters can be expressed in term of $\mathcal{N}_{\rm GR}$ as
\bea\label{epsilon and eta1}
\epsilon_{\rm  v}
                      =    \f{n}{4   \mathcal{N}_{\rm GR}  }  \l( \f{1}{ 1 - 4\kappa \dot{H}}\r)\,,  \qquad
\eta_{\rm v}
             =       \f{n-1}{2   \mathcal{N}_{\rm GR}}  \l(  \f{1}{1 - 4\kappa \dot{H}   } \r)\,, \qquad
\eta_{\k}   =       \f{\k V_0^2  2^{n-2}  M_{\rm P}^{2n-6} }{9  (1 - 4\kappa \dot{H})^3 } (n-2)n^{n+1}  \mathcal{N}_{\rm GR}^{n-2}\,,
\eea
where the $\epsilon_{\rm  v, GR} \equiv  {n}/{(4   \mathcal{N}_{\rm GR})  }   $  and  $  \eta_{\rm v, GR}  \equiv  (n-1)/(2   \mathcal{N}_{\rm GR})$. The spectral index is hence
\be\label{spectral tilt}
n_{\rm s}  \simeq 1  -     \f{n+2}{2   \mathcal{N}_{\rm GR}       }  \l(\f{1}{ 1 - 4\kappa \dot{H} } \r)
-    2^{n-1}  \f{\k V_0^2  M_{\rm P}^{2n-6} }{9 (1-4\kappa \dot{H})^{3} } (n-2)n^{n+1}  \mathcal{N}_{\rm GR}^{n-2}\,.
\ee
For the $V = V_0  \phi^2 $ case, we have
\be \epsilon_{\rm  v}  = \eta_{\rm v} =  \f{1}{2   \mathcal{N}_{\rm GR}  }  \l( \f{1}{ 1 - 4\kappa \dot{H}}\r), \hspace{0.2cm}  \eta_{\k} = 0 \hspace{0.4cm}   \text{and} \hspace{0.4cm}
n_{\rm s}  \, =\,  1 -  \f{2}{  \mathcal{N}_{\rm GR}  (1 - 4\kappa \dot{H})  }\,,   \label{eq_yoyoy}
\ee
and for the $V = V_0 \phi^4 $ case, they are
\be \epsilon_{\rm  v}  =  \f{1}{\mathcal{N}_{\rm GR}  }  \l( \f{1}{ 1 - 4\kappa \dot{H}}\r), \hspace{0.2cm}
\eta_{\rm v} = \f{3}{2\mathcal{N}_{\rm GR}  }  \l( \f{1}{ 1 - 4\kappa \dot{H}}\r),
\hspace{0.2cm}  \eta_{\k} =   \f{8192}{9}  \f{ \k V_0^2 M_{\rm P}^2  \mathcal{N}_{\rm GR}^2}{ (1-4\kappa \dot{H})^3}    \label{eq_yoyoy2}
\ee
and
\be
n_{\rm s}   =  1 -  \f{3}{  \mathcal{N}_{\rm GR}  (1 - 4\kappa \dot{H})    }
-    \f{16384}{9}  \f{\k V_0^2  M_{\rm P}^{2} }{ (1-4\kappa \dot{H})^{3} }\,  \mathcal{N}_{\rm GR}^{2}\,.     \label{eq_otooo}
\ee
The GR predictions for the $V \propto  \phi^n $ models, given $\mathcal{N}_{\rm GR} = 60$ and
$n = 2$, are the tensor-to-scalar ratio $ r \simeq 16\epsilon_{\rm v, GR} \simeq 0.13 $ and $n_{\rm s} \simeq 0.967$. The $n = 4$  case has $ r \simeq 0.27$ and $n_{\rm s} \simeq 0.95$. These are disfavored by Planck 2015's results \cite{Ade:2015lrj} which are
$r < 0.12$ at 95\% CL ($B$-mode polarization constraint from the BICEP2/Keck Array/Planck joint analysis)
and $n_{\rm s} = 0.968 \pm 0.006$ (temperature and large angular scale polarization data).
Eq. (\ref{eq_dotHbaba}) for the power-law potential case is
\be
\dot{H}  \simeq \f{\sqrt{3}  \sqrt{V_0} }{6 } n \phi^{(n-2)/2}  \f{\dot{\phi}  }{ M_{\rm P}  }\,.   \label{eq_opppp}
\ee
For $V = V_0 \phi^2$, it is found from Eq. (\ref{eq_yoyoy}) that the range, $\k \dot{H} \lesssim - 0.027$ can satisfy the upper bound of $r < 0.12$, hence $\k > 0$ is favored.
The Planck 2015 bound of $n_{\rm s} = 0.968 \pm 0.006$ corresponds to the range $-0.071 \lesssim \k \dot{H}   \lesssim  0.031$, hence the combined bound is
 \be 0.071 \; \gtrsim \;   \k |\dot{H}| \;  \gtrsim \;  0.027\,.   \label{eq_thth} \ee
Using $\dot{H} \simeq \sqrt{V_0/3} (\dot{\phi}/M_{\rm P})$   (the $n=2$ case of Eq. (\ref{eq_opppp})), the combined condition (\ref{eq_thth})  is satisfied when
\be
  \f{0.174}{|\dot{\phi}| (m /M_{\rm P})}  \gtrsim \, \k   \, \gtrsim  \,  \f{0.066}{|\dot{\phi}| (m /M_{\rm P})}\,,
\ee
where $V_0 \equiv (1/2) m^2 = \lambda M_{\rm P}^2$ and hence $\lambda = (1/2)(m/M_{\rm P})^2$. The small value of $\dot{\phi}$ and the mass $m > M_{\rm P}$ counterbalance the value of $\k$. For $V = V_0 \phi^4$, using Eq. (\ref{eq_yoyoy2}), the range $\k \dot{H} \lesssim - 0.306$ satisfies the bound of $r < 0.12$. Using $\dot{H} \simeq 2 \sqrt{V_0/3} (\phi \dot{\phi}/M_{\rm P})$   (the $n=4$ case of Eq. (\ref{eq_opppp})), the condition $r < 0.12$ is satisfied as
$
 \k   \,\lesssim  \,  {0.264}/[{ \sqrt{\lambda} \, |\dot{\phi}| (\phi /M_{\rm P})}]
$
where in this case $V_0 = \lambda$.  As seen in Eq. (\ref{eq_otooo}), $n_{\rm s}$ of the $V = V_0 \phi^4$ case needs to be much fine-tuned due to the NMDC-Palatini effect of the $\eta_{\k}$ term. It is noticed that $\k>0$ results in superluminal nature of the metric, less blackhole's entropy and stronger gravitational constant.

\section{Conclusions}  \label{sec_con}
We investigate cosmology of a non-minimal derivative coupling (NMDC) to gravity model in Palatini formalism imposing non-minimal constant coupling between the Einstein tensor and the scalar field derivative term. The Lagrangian contains also a free scalar field derivative term and a scalar potential as proposed in \cite{Sushkov:2009}.
In Palatini formalism, the connection field is a dynamical variable hence Ricci scalar and Ricci tensor are also functions of connection field. As a  result, the Einstein tensor is a function of the connection, $\tilde{G}_{\mu\nu}(\Gamma)$.
Variation of the NMDC action with respect to the independent connection gives the factor, $f \,=\,1 - \kappa{\phi}^{,\alpha}{\phi}_{,\alpha}/2 $. In FLRW spacetime the  factor takes the form,
$f({\dot{\phi}})=1+ {\kappa}{\dot{\phi}}^2/2$. The NMDC coupling constant is enforced to be in a range
of $ -2/ \dot{\phi}^{2} <   \kappa             \; \leq\; \infty $ in order to preserve the
 Lorentz signature of the effective metric. The coupling needs to be negative in order to prevent graviton traveling with superluminal speed. The effective gravitational coupling of the theory is
  $
G_{\rm eff} = {(8 \pi)^{-1}}{\big(1+ {\kappa}{\dot{\phi}}^2/2 \big)^{2}} $ (in the unit of $8 \pi G_{\rm N} = c = 1$)
which reduces to standard GR case when there is no NMDC coupling.
The NMDC-Palatini effective gravitational constant leads to modification of the entropy of blackhole's apparent horizon to
$ S_{\rm AH} = {A}/[{4 (1+\frac{\kappa}{2}{\dot{\phi}}^2)^2}/ 8 \pi].$  The cosmological field equations found can reduce to standard form in the GR limit.
Field equations are approximated in the slow-roll regime. We see that the acceleration condition is modified to $\; w_{\rm eff} \lesssim
- (1/3) \big( 1 +   2\kappa \dot{\phi}^2 \big) $. The NMDC-Palatini effect of the $2 \kappa \dot{\phi}^2$ term, with $\k > 0$, results in acceleration to occur at $w_{\phi}$ value less than $-1/3$. The $\k > 0$ case can also enhance the amount of inflation.
The NMDC-Palatini effect results in an extra term $\eta_{\k}$ in the slow-roll parameter $  \delta   =  - \eta_{\rm v} + \epsilon_{\rm v} + \eta_{\k}$ so that it affects the spectrum index.
 In case of $V \propto \phi^2$, the $\eta_{\k}$ term vanishes in the expression of $n_{\rm s}$, however in the case of $V \propto \phi^4$, it contributes to enormous value of $n_{\rm s}$. Therefore, in this model, the quartic potential is not likely to be viable compared to the Planck 2015's predicted range of spectrum index  \cite{Ade:2015lrj}.
 The $\k > 0$ case can help avoiding super-Planckian region so that it can achieve slow-roll.
 In the $V \propto \phi^2$ case, the $\k > 0$ coupling could help resolving the Planck 2015's tensor-to-scalar ratio constraint, $r < 0.12$. In the GR case, for $\mathcal{N}_{\rm GR} = 60$, the $V \propto \phi^2$ potential gives
$ r  \simeq 0.13 $ and $n_{\rm s} \simeq 0.967$ which are disfavored by Planck 2015. The NMDC-Palatini model, the range $ 0.071 \gtrsim \k |\dot{H}|   \gtrsim   0.027$ can satisfy Planck 2015 tensor-to-scalar ratio upper bound ($r < 0.12$) and the constraint $n_{\rm s} = 0.968 \pm 0.006$ at the same time. The viable range corresponds to
$
  \f{0.174}{|\dot{\phi}| (m /M_{\rm P})}  \gtrsim  \k    \gtrsim  \,  \f{0.066}{|\dot{\phi}| (m /M_{\rm P})},
$
of which  the chaotic inflation $V \propto \phi^2$ could be viable in the range. It should be noted that the positive $\k$ would allow superluminal nature of the metric, less blackhole's entropy and stronger gravitational constant. The other types of inflationary potential such as exponential potential and cosmological perturbations should be investigated in future works. Moreover, analysis on inflationary exits and possibility that the model could give eternal inflation are await to be done.

\section*{Acknowledgements}
This project is supported by Naresuan University Research Grant (R2557C122).
 The authors thank
Antonio De Felice, Phongsaphat Rangdee, Pichet Vanichchapongjaroen and Phongpichit Channuie for discussions.

\appendix
\section{Proof of relation between $\tilde{T}_{\mu\nu}$ and $T_{\mu\nu}$} \label{sec_ap}
\bea\label{energy momentumtensor two frame}
\tilde{T}_{\mu\nu} &=&  -\frac{2}{\sqrt{-h}}\frac{\delta \mathcal{L}_{\rm m}(g_{\kappa\lambda},\Psi)}{\delta h^{\mu\nu}}
= -\frac{2}{f^{2}\sqrt{-g}}\frac{\delta \mathcal{L}_{\rm m}(g_{\kappa\lambda},\Psi)}{\delta (f^{-1} g^{\mu \nu})}
\no \\&=&
-\frac{2}{f\sqrt{-g}} \frac{\delta \mathcal{L}_{\rm m}(g_{\kappa\lambda},\Psi)}{\delta  g^{\mu \nu}} \,  = \, f^{-1} T_{\mu\nu}\,.
\eea
\section{Hubble parameter and the Friedmann equation derived with the metric $h_{\mu\nu}$} \label{sec_hmetricFR}
As in Eq. (\ref{new metric NMDC}), the line element in the new metric $h_{\mu\nu}$ form is $\d s^2 = h_{\mu\nu} \d x^{\mu}\d x^{\nu} = -f \d t^2  + f a^2 \d \bf{x}^2 $. Defining $\d \tilde{t}  = \sqrt{f} \d t$, $\tilde{a}= \sqrt{f} a $ and $\tilde{H} = \tilde{a}^{-1}\d{\tilde{a}}/\d \tilde{t}$, hence $\d s^2 = -\d \tilde{t}^2  +  \tilde{a}^2 \d \bf{x}^2 $. Therefore
\bea
H \,=\,  {\tilde{H}}{\sqrt{f}}  - \f{1}{2}\f{\dot{f}}{f}\,,
\eea
and the Friedmann equation of the $h_{\mu\nu}$ is
\bea
\tilde{H}^2 \,=\,  \f{H^2}{(1+ \kappa \dot{\phi}^2/2)} + \f{\kappa \dot{\phi} \ddot{\phi} H}{(1+ \kappa \dot{\phi}^2/2)^2} + \f{\kappa^2 \dot{\phi}^2 \ddot{\phi}^2 }{4 (1+ \kappa \dot{\phi}^2/2)^3}.
\eea
In the slow-roll regime, $
\tilde{H}^2 \, \simeq \,  {H^2}/{(1+ \kappa \dot{\phi}^2/2)}.
$

\section{Conclusion of results in comparison to the metric NMDC gravity} \label{sec_mp}
Here we compare major results of the metric and Palatini approaches of the NMDC models. Please note that our coupling $\kappa$ is defined as $M^{-2}$ in some other Refs. such as in \cite{Germani:2009, Germani:2010gm, Koutsoumbas:2013boa, Shinji:2015}.

\begin{itemize}
\item {\bf Friedmann equation}\\
\underline{Metric approach} \cite{Sushkov:2009} \\
\be
H^2 \:=\: \f{1}{3 M_{\rm P}^2}\l( {\rho}_{\rm tot}  -  \f{9 \kappa H^2 \dot{\phi}^2}{2}  \r)\,.
\ee
\underline{Palatini approach} \\
From Eq. (\ref{Friedmann slowroll}),
\be
H^2 \;
   \simeq \;  \f {{\rho}_{\rm tot}}{3 M_{\rm P}^2}
  \l( 1+\frac{3}{2}\f{\kappa \dot{\phi}^2}{M_{\rm P}^2}(1+w_{\rm eff} )\r)\,,
\ee
where
$\rho_{\rm tot}  = {\varepsilon \dot{\phi}^2}/{2} + V(\phi) +  \rho_{\rm m}$. Of the metric approach, there is a NMDC coupling term $ \kappa H^2 \dot{\phi}^2$ in the Friedmann equation. This is coupled to kinematic part, the Hubble function.
Unlike the metric case, in the NMDC Palatini Friedmann equation, the NMDC coupling to the kinematic part is not via the $H$  or $\dot{H}$  terms but via the effective EoS, $w_{\rm eff}$ which is either written as function of density or function of the $H$  or $\dot{H}$ terms.
\item {\bf Klein-Gordon equation}\\
\underline{Metric approach}   (See, e.g. \cite{Sadjadi:2010bz, Gumjudpai:2015vio})
\be
\ddot{\phi}\l(\varepsilon - 3 \kappa H^2  \r) +
3 H \dot{\phi}\l(   \varepsilon -  3 \kappa H^2 - 2 \kappa \dot{H}    \r)  + V' =  0\,.
\ee
\underline{Palatini approach} \\
Eq. (\ref{eq_KGpalaq}) reads,
\be
\ddot{\phi}\l( \varepsilon - (9/2) \kappa \dot{H}  -  {(15/2) \kappa}H^2 \r)  +  3H \dot{\phi} \l( \varepsilon - 4 \kappa\dot{H}  \r)   + V'  \;  \simeq \;  0\,.  \nonumber
\ee
\item {\bf Spectral index}\\
\underline{Metric approach} \\ During slow-roll, $H^2 \gg \dot{H}$ and  in high friction limit $(-\kappa H^2 \gg 1)$     \cite{Germani:2011ua, Dalianis:2016wpu},
\be
1-n_{\rm s}  \: \simeq \:  \f{ 8 \epsilon_{\rm v, GR}}{(\varepsilon - 3 \kappa H^2)}  -   \f{2 \eta_{\rm v, GR}}{(\varepsilon - 3 \kappa H^2)}\,.
\ee
\underline{Palatini approach}\\
From Eq. (\ref{eq_nspala}), in high friction limit, $- \kappa\dot{H} \gg 1 $,
\be
1-n_{\rm s}  \: \simeq \:  - \f{6 \epsilon_{\rm v, GR}}{(\varepsilon - 4 \kappa \dot{H})}  +   \f{2 \eta_{\rm v, GR}}{(\varepsilon - 4 \kappa \dot{H})}\,,
\ee
where $\epsilon_{\rm v, GR} \equiv  (M_{\rm P}^2 /2)(V'/V)^2$ and $\eta_{\rm v, GR} \equiv  M_{\rm P}^2(V''/V)$.

\end{itemize}

\end{document}